\documentclass[a4paper,12pt]{article}
\usepackage{literat}
\usepackage{geometry}
\usepackage{graphicx}

\title{HOPF BIFURCATION WITHIN THERMODYNAMIC REPRESENTATION}

\author{A.I. Olemskoi\footnote{E-mail: alex@ufn.ru}, I.A. Shuda}

\begin{document}
\maketitle
\begin{center}
Sumy State University\\
Rimskii-Korsakov St. 2, 40007 Sumy, Ukraine
\end{center}

\begin{abstract}
 On base of Hamiltonian formalism, we show that Hopf bifurcation arrives, in
the course of the system evolution, at creation of revolving region of the
phase plane being bounded by limit cycle. A revolving phase plane with a set of
limit cycles is presented in analogy with revolving vessel containing
superfluid He$^4$. Within such a representation, fast varying angle is shown to
be reduced to phase of complex order parameter whose module squared plays a
role of action. Respectively, vector potential of conjugate field is reduced to
relative velocity of movement of the limit cycle interior with respect to its
exterior.

{\bf Key words:} Fast and slow variables; Limit cycle; Gauge field; Order
parameter.

PACS numbers: 05.45.-a, 05.65.+b
\end{abstract}

\section{Introduction}

It is not gross exaggeration to assert that conception of the phase transition
is one of fundamental ideas of contemporary physics. Related picture is based
on the Landau scheme, according to which a thermodynamic system, driven by slow
and monotonic variations of state parameters of heat bath, rebuilds its
macroscopic state if thermodynamic potential gets one or more additional minima
in a state space \cite{L}. From mathematical point of view, such a phase
transition represents the simplest bifurcation that results in doubling
thermodynamic steady states.

As is known, thermodynamic phase transition is a special case of
self-organization process in the course of which three principle parameters,
being order parameter, its conjugate field and control parameter, very in
self-consistent manner \cite{OC}. Roughly speaking, a generalization of
thermodynamic picture due to passage to synergetic one is stipulated by
extension of set of state parameters from single parameter to three ones, being
above pointed out. It might hope that such a generalization allows one to
describe not only the simplest Landau-like bifurcation, but much more
complicated Hopf one, when a limit cycle is created to be continuous manifold
instead of discrete one \cite{H}.

Our considerations of this problem have shown \cite{UFJ} that using the whole
set of universal deformations within standard synergetic scheme does not arrive
at stable limit cycle, whereas running out off the standard scheme of
self-organization has allowed us to obtain the limit cycle shown in
Fig.1 \cite{1}.
\begin{figure}[!h]
\centering
\includegraphics[width=120mm]{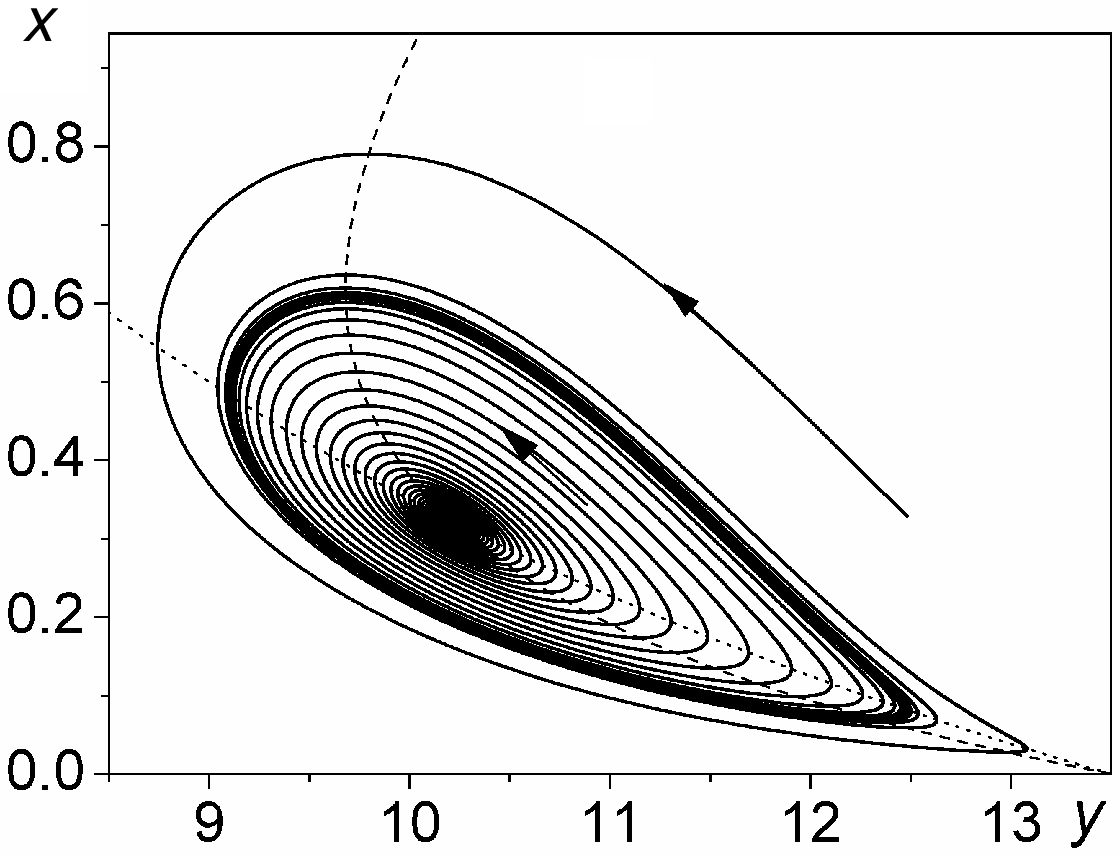}
\caption{The limit cycle related to non-standard self-organization equations
$\dot x = x\left[y-(1+rx)-(A-1)/(1+\rho x)\right]$, $\dot y=A-y(1+x)$ with
$A=14$, $r=5$, $\rho=2$ \cite{1}.}\label{fig.1}
\end{figure}
In this connection, the question arises: what is the physical reason that
self-organization scheme may not represent the Hopf bifurcation?

This paper is devoted to the answering above question. It is appeared, main
reason is that a description of a limit cycle demands using both potential and
force of a field conjugated to an order parameter, whereas standard synergetic
scheme uses an force of this field only. Following \cite{2}, we show in Section
2 that fast revolving of the state point in the phase plane induces a gauge
field, whose potential is reduced to relative velocity of movement of interior
domain of the limit cycle with respect to the exterior. Such a picture allows
us to study, in Section 3, a revolving phase plane with a set of limit cycles,
using an analogy with revolving vessel containing superfluid He$^4$ \cite{9}.
Section 4 concludes our consideration.

\section{Hopf bifurcation within canonical representation}

We consider Hamiltonian dynamics determined by the equations of motion
\begin{equation}
\dot {q}_i=\frac{\partial H}{\partial p_i},\quad \dot{p}_i=-\frac{\partial
H}{\partial q_i};\qquad \{{q}_i\}=q,Q,\ \ \{{p}_i\}=p,P
 \label{0}
\end{equation}
for both fast and slow coordinates $q$, $Q$ and conjugate momenta $p$, $P$,
respectively (hereafter, the dot over a symbol denotes time derivation).
Related Hamiltonian
\begin{equation}
H(q,p;Q,P)=H_s(Q,P)+H_f(q,p;Q)
 \label{1}
\end{equation}
is splitted into the slow term $H_s(Q,P)$ and the fast one $H_f(p,q;Q)$,
latter of which depends on the slow coordinate also.

In accordance with standard scheme \cite{LL} it is naturally to pass from the
fast variables $q$, $p$ to canonical ones, being fast alternating angle
$\varphi$ and slow varying action $\eta^2$. This passage keeps invariant the
first term of the Hamiltonian (\ref{1}) and transforms the second one according
to relation
\begin{equation}
H^{'}_f(\varphi;\eta,Q)=H_f(q,p;Q)+\dot Q\frac{\partial\Psi(q;\eta,Q)}{\partial
Q}
 \label{3}
\end{equation}
whose explicit form is determined with generating function $\Psi(q;\eta,Q)$ to
be defined by the following constrains:
\begin{equation} \frac{\partial\Psi(q;\eta,Q)}{\partial
q}=p,\quad \frac{\partial \Psi(q;\eta,Q)}{\partial Q}=P,\quad \frac{\partial
\Psi(q;\eta,Q)}{\partial\eta^2}=\varphi.
 \label{2}
\end{equation}
Due to fast variations of the angle $\varphi$, it is naturally to consider the
term (\ref{3})
\begin{equation}
H^{'}(\eta,Q)\equiv\langle
H^{'}_f(\varphi;\eta,Q)\rangle\equiv\frac{1}{2\pi}\int\limits_0^{2\pi}H^{'}_f(\varphi;\eta,Q){\rm
d}\varphi,
 \label{4}
\end{equation}
being averaged over this variations.

To find related Hamiltonian one has to use one-valued generating function
\begin{equation}
\Phi(\varphi;\eta,Q)\equiv\Psi\Big(q(\varphi;\eta,Q);\eta,Q\Big),\quad
0\leq\varphi\leq 2\pi
 \label{5}
\end{equation}
instead of many-valued one, $\Psi(q;\eta,Q)$. Then, the last factor in Eq.
(\ref{3}) is determined by the relation
\begin{equation}
\frac{\partial\Phi}{\partial Q}=\frac{\partial\Psi}{\partial Q}+p\frac{\partial
q}{\partial Q}
 \label{6}
\end{equation}
that expresses the chain rule. Its using gives the averaged term (\ref{4}) in
the following form:
\begin{equation}
H^{'}(\eta,Q)=H(\eta,Q)+\dot Q\left<\frac{\partial\Phi}{\partial
Q}-p\frac{\partial q}{\partial Q}\right>,\quad H(\eta,Q)\equiv\langle
H_f(q,p;Q)\rangle.
 \label{7}
\end{equation}

As a result, usage of the canonical angle-action representation derives to the
transformation of the averaged Hamiltonian (\ref{1}):
\begin{eqnarray}
H_{ef}(\eta;Q,P)&\equiv &\langle H^{'}(\varphi,\eta;P,Q)\rangle\nonumber\\
&=&\mathcal{H}(\eta;Q,P)+\dot Q\left(\left<\frac{\partial\Phi}{\partial
Q}\right>-\left<p\frac{\partial q}{\partial Q}\right>\right)
 \label{8}
\end{eqnarray}
where the first term
\begin{equation}
\mathcal{H}(\eta;Q,P)\equiv H(\eta,Q)+H_s(Q,P)
 \label{9}
\end{equation}
depends on slow varying values only.

To find equations of motions for above slow variables one has to issue from the
extremum condition for effective action \cite{LL}
\begin{equation}
S_{ef}\{Q(t),P(t);\eta(t)\}\equiv\int_{t_{in}}^{t_{f}}\left[P(t)\dot Q(t)-
H_{ef}\Big(\eta(t);Q(t),P(t)\Big)\right]{\rm d}t
 \label{10}
\end{equation}
whose form is determined by the Hamiltonian (\ref{8}); $t_{in}$, $t_{f}$ are
initial and final points of the time. Variation of this expression with respect
to the momentum derives to the equation of motion
\begin{equation}
\dot Q_\alpha =\frac{\partial\mathcal{H}}{\partial P_\alpha}
 \label{11}
\end{equation}
keeping initial form (\ref{0}) (hereafter, we suppose that slow variables are
vector quantities with components $Q_\alpha$, $P_\alpha$). On the other hand,
variation of the action (\ref{10}) over the slow coordinate arrives at the
equation
\begin{equation}
\dot P_\alpha=-\frac{\partial\mathcal{H}}{\partial
Q_\alpha}+F_{\alpha\beta}\frac{\partial\mathcal{H}}{\partial P_\beta}
 \label{12}
\end{equation}
that is prolonged due to an effective field with the force given by
antisymmetric tensor
\begin{equation}
F_{\alpha\beta}\equiv\frac{\partial A_\beta}{\partial Q_\alpha}-\frac{\partial
A_\alpha}{\partial Q_\beta}
 \label{13}
\end{equation}
and vector potential
\begin{equation}
A_\alpha\equiv\left<p_\beta\frac{\partial q_\beta}{\partial Q_\alpha}\right>.
 \label{14}
\end{equation}
As usually, we use the Einstein summation rule over repeated index $\beta$.

Above relations (\ref{12})--(\ref{14}) arrive at the conclusion of fundamental
importance: fast variations of coordinates whose values depend on slow
variables induces an effective gauge field \cite{2}. For case of the Hopf
bifurcation, it means that such a bifurcation, in the course of the system
evolution, arrives at the revolving not only of the configuration point, but of
the whole domain of the phase plane that is bounded by the limit cycle. In
other words, the physical picture of the Hopf bifurcation means the phase plane
behaves as a real object, but not a mathematical one.

\section{Physical picture of limit cycles}

We consider a round phase plane that is spanned on the axes of both coordinate
$q$ and momentum $p$ and revolves with the angle velocity $\vec\omega_0$ and a
moment of inertia $I$. From the physical point of view, the value
$\vec\omega_0$ determines the frequency of external influence, whereas the
quantity $I$ is reduced to total action of the system under consideration. If
the phase circle revolves as a solid plane, a phase point with coordinate $\bf
r$ has the linear velocity ${\bf v}_n=[\vec\omega_0,{\bf r}]$.

According to above consideration, the Hopf bifurcation results in the limit
cycle creation that induces a gauge field with the vector potential (\ref{14})
and the strength (\ref{13}), which are reduced to linear and angle velocities,
${\bf w}$ and $\vec\omega$, respectively. These are not equal to normal values
${\bf v}_n$ and $\vec\omega_0$ because a region bounded by the limit cycle
revolves with different velocities due to the gauge field effect. Indeed, if
one represents the limit cycle creation as an ordering with a complex parameter
$\phi=\eta{\rm e}^{{\rm i}\varphi}$, then the phase gradient ${\bf v}_s\equiv
s\nabla\varphi$, where $\nabla\equiv\partial/\partial{\bf r}$, $s$ being an
elementary action, affects in such a manner to compensate a rotation within a
domain bounded by the limit cycle: ${\bf w}={\bf v}_n-{\bf v}_s$. In this way,
the relative velocity ${\bf w}$ appears as a gradient prolongation:
$\nabla\Rightarrow\nabla-({\rm i}/s){\bf w}$, being caused by the gauge field
${\bf w}$. In opposition to the case of the solid plane revolving, ordering
arrives at non-linear relation $\vec\omega=(1/2){\rm rot}~{\bf w}$ between the
angular and linear components of the revolving velocity.

Well-known example of such type behaviour represents the case of revolving
superfluid He$^2$ \cite{9}. Along this line, an effective potential density of
the revolving phase plane, including a set of limit cycles, has the following
form \cite{GLA}
\begin{equation}
E=\Delta E(\eta)+\frac{1}{2}\Big|\left(-{\rm i}s\nabla-{\bf
w}\right)\eta\Big|^2+\frac{I}{2}\omega^2.
 \label{15}
\end{equation}
Within the phenomenological scheme, the density of the potential variation due
to limit cycle creation is given by the Landau expansion
\begin{equation}
\Delta E(\eta)=A\eta^2+\frac{B}{2}\eta^4
 \label{16}
\end{equation}
whose form is fixed by parameters $A$, $B$. The second term of Eq.(\ref{15})
determines heterogeneity energy with the gradient, being prolonged by the
vector potential ${\bf w}$ of the gauge field. The last term is the kinetic
energy of the revolving phase plane.

Under an external influence with frequency $\vec\omega_0$, the system behaviour
is defined by the effective potential density
\begin{equation}
\widetilde E= E-\left(I\vec\omega_0+{\bf M}\right)\vec\omega_0
 \label{17}
\end{equation}
whose value is determined with respect to the revolving plane that is
characterized with the angular momentum ${\bf
M}=I\left(\vec\omega-\vec\omega_0\right)$. Steady state distributions of the
order parameter $\eta(\bf r)$ and the relative velocity ${\bf w}(\bf r)$ are
given by extremum condition of the total value of the effective potential
\begin{equation}
\mathcal{E}\{\eta({\bf r}),{\bf w}({\bf r})\}=\int\widetilde E\Big(\eta(\bf
r),{\bf w}(\bf r)\Big){\rm d}\bf r
 \label{18}
\end{equation}
where integration is fulfilled over the whole area of the phase plane. In this
way, the boundary conditions are as follows:
\begin{itemize}
\item out off a limit cycle
\begin{equation}
\eta=0,\ \ \nabla\eta=0,\ \ {\bf w}=[\vec\omega_0,{\bf r}], \ \
\vec\omega=\vec\omega_0;
 \label{19}
\end{equation}
\item within a limit cycle
\begin{equation}
\eta=\eta_0,\ \ \nabla\eta=0,\ \ {\bf w}=0, \ \ \vec\omega=0;
 \label{20}
\end{equation}
\item on a limit cycle itself
\begin{equation}
{\bf n}\left(-{\rm i}s\nabla-{\bf w}\right)\eta=0.
 \label{21}
\end{equation}
\end{itemize}
Here, ${\bf n}$ is the unit vector being perpendicular to the limit cycle,
$\eta_0=\sqrt{-A/B}$ is the stationary value of the order parameter to be
determined by the minimum condition of the expression (\ref{16}).

According above expressions, a disordered phase related to exterior of the
limit cycle is characterized with the effective energy density $\widetilde
E(0)=-(I/2)\vec\omega_0^2$, whereas an ordered phase being bounded by this
cycle relates to the value $\widetilde E(\eta_0)=-(|A|/2)\eta_0^2$. As a
result, the condition $\widetilde E(\eta_0)=\widetilde E(0)$ of the phase
equilibrium gives a characteristic value of the revolving velocity
\begin{equation}
\omega_c\equiv\sqrt{\frac{|A|\eta_0^2}{I}}=\sqrt{\frac{A^2}{IB}}
 \label{22}
\end{equation}
to determine an energy scale $E_c\equiv I\omega_c^2=|A|\eta_0^2=A^2/B$.
Moreover, it is useful to introduce two lengths $\lambda$, $\xi$ and their
ratio $\kappa=\lambda/\xi$ to be determined by the following relations:
\begin{equation}
\lambda\equiv\sqrt{\frac{IB}{4|A|}},\quad \xi\equiv
\sqrt{\frac{s^2}{2|A|}};\qquad\kappa\equiv\sqrt{\frac{I}{I_0}}, \quad I_0\equiv
\frac{2s^2}{B}.
 \label{23}
\end{equation}
Then, measuring the energy density $\widetilde E$ in units $E_c$, the order
parameter $\eta$ -- in $\eta_0$, the angle velocity $\vec\omega$ -- in
$\omega_c$, the linear velocities ${\bf v}_n$, ${\bf w}$ -- in
$2\sqrt{2}\lambda\omega_c$, the angular moment ${\bf M}$ -- in
$2\sqrt{2}I\lambda\omega_c$, and the distance $r$ -- in $\lambda$, one reduces
the energy density (\ref{17}) to the simplest form
\begin{equation}
\widetilde E=\Big|\left(-{\rm i}\kappa^{-1}\nabla-{\bf
w}\right)\eta\Big|^2-\left(\eta^2-\frac{1}{2}\eta^4\right)-\left(\vec\omega_0-
\frac{1}{2}\vec\omega\right)\vec\omega.
 \label{24}
\end{equation}
Inserting this equality into the total energy (\ref{18}) and variating the
functional obtained, one finds the following equations of motion:
\begin{equation}
\kappa^{-2}\nabla^2\eta=-\left(1-{\bf w}^2\right)\eta+\eta^3,
 \label{25a}
\end{equation}
\begin{equation}
-{\rm rot}~{\rm rot}~{\bf w}=\eta^2{\bf w}.
 \label{25b}
\end{equation}

As is known \cite{9}, the form of solutions of these equations is fixed with
the parameter $\kappa$ given by two last Eqs.(\ref{23}). In usual case, the
phase plane is so small to be realized the condition $\kappa\leq 2^{-1/2}$, and
a single limit cycle (type of shown in Fig.\ref{fig.1}) can be created with the
form and size determined by the external frequency $\omega_0$. Much more rich
situation is realized in the case of the so large phase plane that inverted
condition $\kappa>2^{-1/2}$ is fulfilled. Then, within the interval
$\omega_{c1} <\omega_0< \omega_{c2}$, bounded with the limit velocities
\begin{equation}
\omega_{c1}\equiv\frac{\ln\kappa}{\sqrt{2}\kappa}\omega_{c}=\frac{|A|}{4s}
\left(\frac{I}{I_0}\right)^{-1}\ln\frac{I}{I_0},
 \label{26}
\end{equation}
\begin{equation}
\omega_{c2}\equiv\sqrt{2}\kappa\omega_{c}=|A|/s,
 \label{27}
\end{equation}
the mixed state is realized to be a set of round limit cycles periodically
distributed over surface of the revolving phase plane. Each of these cycles has
elementary action $2\pi s$ to reach the maximum value $N_{\max}=1/\pi\xi^2$ of
the cycle density per unit of area at $\omega_0=\omega_{c2}$. With falling down
external velocity near the upper boundary
$(0<\omega_{c2}-\omega_0\ll\omega_{c2})$ the limit cycle density decreases
according to the equality
\begin{equation}
{N\over N_{\max}}={\omega_0\over\kappa}- {\overline{\eta^2}\over 2\kappa^2}
 \label{28}
\end{equation}
where average over the phase plane $\overline{\eta^2}$ is connected with the
revolving velocity $\omega_0$ by the equality
\begin{equation}
\overline{\eta^2}={2\kappa\over\beta(2\kappa^2-1)} (\kappa-\omega_0),\ \ \beta
\equiv \overline{\eta^4}/(\overline{\eta^2})^2=0.1596.
 \label{29}
\end{equation}
The average value
\begin{equation}
\overline{\omega}=\omega_0-\overline{\eta^2}/2\kappa=
\omega_0-(\kappa-\omega_0)/\beta(2\kappa^2-1)
 \label{30}
\end{equation}
is smaller than external value $\omega_0$ in a quantity being equal to the
average of the plane polarization
\begin{equation}
\overline{M}=-\overline{\eta^2}/2\kappa =
-(\kappa-\omega_0)/\beta(2\kappa^2-1).
 \label{31}
\end{equation}
The maximum value of a revolving velocity is reached in cores of limit cycles,
and the minimum one $\omega_{\min}=\omega_0-\sqrt{2}(\kappa-\omega_0)/
(2\kappa^2-1)$ -- in the centers of triangles formed by cycles. The average
variation of effective energy (\ref{18}) caused by the phase plane revolving
\begin{equation}
\overline{\mathcal E}=I\omega^2_c\left({1\over 2}+\overline{\omega^2}-
{\overline{\eta^4}\over 2}\right)= I\omega^2_c\left[ {1\over
2}+\overline{\omega}^2- {(\kappa-\overline{\omega})^2\over
1+\beta(2\kappa^2-1)}\right]
 \label{32}
\end{equation} is the function of
the average velocity $\overline{\omega}$, differentiation with respect to which
results in Eq.(\ref{30}).

Near the lower critical value $\omega_{c1}$, the limit cycle density
$N=(\kappa/2\pi)\overline{\omega}$ is not so large and these cycles can be
treated independently. Taking into account that $w(r)$ varies at distances
$r\sim 1$ and $\eta(r)$ does at $r\sim \kappa^{-1}\ll 1$, the relative velocity
are determined by Eq.(\ref{25b}) with $\eta^2\approx 1$ and $\kappa\gg 1$:
\begin{equation}
w=-\kappa^{-1}K_1(r) \label{33}
\end{equation}
where $K_1({\bf r})$ is the Hankel function of imaginary argument.
Respectively, the order parameter is determined by Eq.(\ref{25a}) with
$w=-1/\kappa r$:
\begin{eqnarray}
\eta\simeq cr\quad {\rm at}\quad r\ll\kappa^{-1},\nonumber\\ \eta^2 \simeq
1-(\kappa r)^{-2}\quad {\rm at}\quad r\gg\kappa^{-1} \label{34}
\end{eqnarray}
where $c$ is positive constant. According to Eq.(\ref{33}) one has $w\approx
-1/\kappa r$ at $r\ll 1$ and $w\approx -\sqrt{\pi/2\kappa^2}~r^{-1/2}e^{-r}$ at
$r\gg 1$. The dependence $\overline{\omega}(\omega_0)$ is of steadily
increasing nature: at $\omega_0=\omega_{c1}$ it has the vertical tangent and
with $\omega_0$ growth it asymptotically approaches to the straight line
$\overline{\omega} =\omega_0$. Effective energy per one limit cycle is
$(2\pi/\kappa^2)\ln\kappa$, the $\omega$ value in a cycle center is twice
as large as $\omega_{c1}$.

\section{Conclusions}

Within Hamiltonian formalism, combined consideration of both fast and slow sets
of dynamical variables shows that averaging over the angle of the canonical
pair angle-action induces an effective gauge field if fast coordinates depend
on slow ones. For case of the Hopf bifurcation, it means that such a
bifurcation, in the course of the system evolution, arrives at the revolving
not only of the configuration point, but the whole region of the phase plane
being bounded by the limit cycle. In other words, the physical picture of the
Hopf bifurcation means the phase plane behaves as a real object, but not
mathematical one.

Along this line, a revolving phase plane with a set of limit cycles can be
presented in analogy with revolving vessel containing superfluid He$^4$. Within
framework of such a representation, fast varying angle is reduced to the phase
$\varphi$ of the complex order parameter $\phi=\eta{\rm e}^{{\rm i}\varphi}$
whose module squared $\eta^2$ plays a role of the action. In this way, a role
of the vector potential of the gauge field plays the relative velocity $\bf w$
of movement of interior region of the limit cycle with respect to its exterior,
whereas the field force is reduced to the related angle velocity
$\vec\omega=(1/2){\rm rot}~{\bf w}$. By this, slow variables are reduced to the
parameters $A$, $B$ of the Landau expansion (\ref{16}).

Finally, we focus upon the physical interpretation of above picture of a phase
plane with one or more limit cycles. The usual case, when canonical pair of
coordinate and momentum can be varied within the whole area of the phase plane,
relates to the parameter $\kappa\sim 1$, where the Hopf bifurcation arrives at
creation of single limit cycle, which presents periodical variations in the
self-organized system. In opposite case, when a domain of the coordinate and
momentum variations within a limit cycle is much less than the phase plane
area, the principle parameter is $\kappa\gg 1$. If the self-organized system is
subjected to an external periodical influence whose frequencies are bounded
within domain $\omega_{c1}\div\omega_{c2}$ given by Eqs. (\ref{26}),
(\ref{27}), a set of non-trivial resonances is appeared. Mean values of the
coordinate and momentum variations within these resonances relate to the
centers of the limit cycles, whereas variations amplitude is determined by the
correlation length $\xi$ given by the second of definitions (\ref{23}). The
first of such resonances appears at external frequency $\omega_{0}=\omega_{c1}$
to arrive at growth of its number to maximum value $N_{\rm max}\sim\xi^{-2}$ at
the upper frequency $\omega_{0}=\omega_{c2}$. In this way, the resonances
number varies in accordance with Eq.(\ref{28}), whereas average resonance
frequency is defined by Eq.(\ref{30}) which shows its monotonic
increase with external frequency.

\end{document}